\documentclass[aip,jcp,twocolumn,showpacs]{revtex4}
\usepackage{amsfonts,amsmath,amssymb,epsfig}

\usepackage{color}

\newcommand{\la}{\langle}
\newcommand{\ra}{\rangle}
\newcommand{\tr}{{\mathrm{Tr}}}
\newcommand{\ffcf}{{\mathrm{FFCF}}} 
\newcommand{\Ccl}{{\mathcal{C}_{\text{CL}}}} 
\newcommand{\Cqm}{{\mathcal{C}_{\text{QM}}}} 
\newcommand{\Ccan}{{\overline{\mathcal{C}}_{\mathrm{QM}}}} 
\newcommand{\ud}{\mathrm{d}}
\newcommand{\ui}{\mathrm{i}}
\newcommand{\ue}{\mathrm{e}}
\newcommand{\vz}{{\bf z}}
\newcommand{\vp}{{\bf p}}
\newcommand{\vq}{{\bf q}}
\newcommand{\vZ}{{\bf Z}}
\newcommand{\vJ}{{\bf J}}

\newcommand{\todo}[1]{}
\newcommand{\rem}[1]{ }

\begin{document}

\title{The Flux-Flux Correlation Function for Anharmonic  Barriers}

\author{Arseni Goussev$^1$, Roman Schubert$^1$, Holger
  Waalkens$^{2}$, and Stephen Wiggins$^1$}

\affiliation{$^1$School of Mathematics,
  University of Bristol, University Walk, Bristol BS8 1TW, UK\\
  $^2$Johann Bernoulli Institute for Mathematics and Computer Science, University of
  Groningen, PO Box 407, 9700 AK Groningen, The Netherlands}

\date{\today}

\begin{abstract}

  The flux-flux correlation function formalism is a standard and widely used approach for the computation of reaction rates.  In
  this paper we introduce a method to compute the classical and
  quantum flux-flux correlation functions for anharmonic barriers
  essentially analytically through the use of the classical and quantum
  normal forms.  In the quantum case we show that the quantum normal form reduces the computation of the flux-flux correlation function to  that of an effective one dimensional anharmonic barrier. 
  The example of the computation of the quantum flux-flux correlation function for a fourth order anharmonic barrier  is worked out in detail, and we present an analytical expression for the quantum mechanical microcanonical flux-flux correlation function.  
  We then give a discussion of the short-time and
  harmonic limits. 
  \end{abstract}

\pacs{02., 05., 34.10.+x, 34.50.Lf, 82.20.-w, 82.20.Db}

\maketitle

\section{Introduction}

In spite of the tremendous  increase in computer power over the last decades,  the computation of quantum reaction rates still is a formidable task. This is the topic of this paper,  and we begin by giving the setting that is relevant to our work.

The microcanonical 
rate constant is given by
\begin{equation} \label{eq:def_k(E)}
k(E) = \big( 2\pi \hbar \rho_{\text{r}}(E) \big)^{-1} N(E)\,,
\end{equation}
where $\rho_{\text{r}}(E)$ is the density of states of reactants and
$N(E)$ is the cumulative reaction probability, which in turn can be formally expressed in terms of the ${\bf S}$ matrix as 
\begin{equation}\label{eq:N(E)_from_S}
N(E) = \sum_J(2J+1) \sum_{n_{\text{p}},n_{\text{r}}} \vert  S_{n_{\text{p}},n_{\text{r}}}(E,J) \vert^2\,.
\end{equation}
Here the inner sum runs over the asymptotic states of reactants and
products which are labeled by $n_{\text{r}}$ and $n_{\text{p}}$,
respectively, and the outer sum covers all values of total angular
momentum $J$.

Though formally correct, the computation of a reaction rate via the ${\bf S}$ matrix is extremely inefficient since the computationally expensive information related to  the state-to-state reactivities embodied in the ${\bf S}$  matrix is ''thrown away'' as a consequence of the averaging embodied in the summations in \eqref{eq:N(E)_from_S}. Motivated by the success of transition state theory (TST) for computing reaction rates using classical mechanics, many researchers have sought a quantum mechanical version of transition state theory.  Recall that the main idea of TST as invented by Eyring, Polanyi and Wigner in the 1930s is to compute classical reaction rates from the flux through a dividing  surface in phase space which separates the phase space region associated with  reactants from the phase space region associated with products.  Assuming the dividing surface to be given by an equation $s(\vz)=0$, where $s$ is a scalar function on the $2f$-dimensional phase space with coordinates 
$\vz=(\vq,\vp)=(q_1,\ldots,q_f,p_1,\ldots,p_f)$ with $s(\vz)<0$ in the reactants region  and 
$s(\vz)>0$ in the products region, the classical microcanonical rate constant  can be written as
\begin{equation} \label{eq:k_CL}
k_{\text{CL}}(E) =   \rho_{\text{r}}(E)^{-1} (2\pi \hbar)^{-f}   \int \ud \vz \: \delta(E-H(\vz)) F(\vz)\,.
\end{equation}
Here $F(\vz)$ is the so called flux factor which is given by
\begin{equation}  \label{eq:flux_factor_classical}
F(\vz ) = \frac{d}{dt} \Theta\big( s(\vz_t) \big) \bigg|_{t=0}  = -\{ H, \Theta(s)\} (\vz)
\end{equation}
with $\Theta$ denoting the Heaviside function, $\{\cdot,\cdot\}$ denoting the Poisson bracket and
 $\vz_t$ denoting the solution of Hamilton's equations at time $t$ with initial conditions $\vz_0=\vz$, i.e. 
$\vz_t=\Phi^t_H(\vz)$, where  $\Phi_H$  is  the Hamiltonian flow  which acts on 
$\vz$ for time $t$.

For the TST computation of the rate constant  to be useful the dividing surface needs to have the property that it is crossed exactly once by reactive trajectories (i.e. trajectories evolving from reactants to products) and not crossed at all by 
nonreactive trajectories. A dividing surface not satisfying this no-recrossing  property leads to an overestimation of the reaction rate. 
The construction of a recrossing free dividing surface has posed a major problem in the development of TST. 
Formally, recrossing  trajectories are eliminated  by multiplying the integrand in \eqref{eq:k_CL}  by 
the projection function
\begin{equation} \label{eq:projection_operator_classical}
P_{\text{r}} (\vz) = \lim_{t\to +\infty} \Theta \big( s (\vz_t) \big) 
\end{equation}
which evaluates to $1$ if the trajectory $\vz_t= \Phi^t_H (\vz )$
evolves to products for $t\to +\infty$, and to 0 otherwise.  It is then
not difficult to see that the this way corrected expression for the
rate constant can be written in the form
\begin{equation} \label{eq:k(E)_classical_flux}
k_{\text{CL}}(E) =  \big(2 \rho_{\text{r}}(E)\big)^{-1}\int_{-\infty}^\infty \ud t\, \Ccl(E,t) \,,
\end{equation}
where $\Ccl(E,t)$ is the flux-flux correlation function (FFCF)
\begin{equation} \label{eq:FFCF_classical}
\Ccl(E,t) = (2\pi \hbar )^{-f} \int  \ud \vz \: \delta\big(E-H(\vz)\big) F(\vz) F(\vz_t) \,.
\end{equation}

The canonical analogues of the microcanonical expression above are easily obtained from replacing the density of states $\delta(E-H(\vz))$ by its canonical 
counterpart $\exp(-\beta H(\vz))$ ($\beta=(k_{\text{B}} T)^{-1}$ denoting the inverse temperature).
Miller, Schwartz and Tromp \cite{MillerSchwartzTromp83} and  Yamamoto \cite{Yamamoto60} took this as a starting point to  express   the quantum mechanical rate constant in a similar fashion. 
For the microcanonical case, this amounts to replacing the phase space functions above by the corresponding operators and the integrals over phase space by the traces of operators.
For the quantum analogue of the rate constant in \eqref{eq:k_CL}, this leads to
\begin{equation} \label{eq:K(E)_quantum}
k_{\text{QM}} (E) = \rho_{\text{r}}(E)^{-1}   \tr \left\{ \delta(E-\hat{H}) \hat{F} \hat{P}_{\text{r}} \right\} \,,
\end{equation}
where $ \rho_{\text{r}}(E)$ now is the quantum mechanical partition function of the reactants (for which we use the same symbol as in the classical case),  the flux factor \eqref{eq:flux_factor_classical} becomes  the operator
\begin{equation} \label{eq:flux_factor_quantum}
\hat{F} = \frac{d}{dt} \left(    e^{i\hat{H}t/\hbar} \Theta(\hat{s}) e^{-i\hat{H}t/\hbar} \right)
  \bigg|_{t=0} = \frac{i}{\hbar} \big[ \hat{H}, \Theta(\hat{s}) \big]
\end{equation}
and  the projection function \eqref{eq:projection_operator_classical} becomes the operator
\begin{equation} \label{eq:projection_operator_quantum}
\hat{P}_{\text{r}} = \lim_{t\rightarrow +\infty} \ue^{\ui \hat{H}t/\hbar} \Theta(\hat{s})  \ue^{- \ui \hat{H}t/\hbar}\,. 
\end{equation}
Similarly to the classical case expression \eqref{eq:K(E)_quantum} for the quantum rate constant can then also be rewritten in terms of a flux-flux correlation function, namely
\begin{equation} \label{eq:k_QM_flux}
k_{\text{QM}} (E) =  \big(2\rho_{\text{r}}(E)\big)^{-1} \int_{-\infty}^\infty \ud t\, \Cqm(E,t) 
\end{equation}
with
\begin{equation} \label{eq:FFCF_quantum}
\Cqm(E,t) = \tr \left\{ \delta(E-\hat{H}) \hat{F}
    e^{i \hat{H} t/ \hbar} \hat{F} e^{-i \hat{H} t/ \hbar} \right\} \,. 
\end{equation}

Clearly, the ability to compute the quantum mechanical flux-flux correlation function is essential for computing quantum mechanical reaction rates using this approach. 
Exact expressions for the \emph{canonical} quantum mechanical flux-flux correlation function can be found in \cite{MillerSchwartzTromp83} for the free particle  and 
in \cite{Miller74, MillerSchwartzTromp83, WangSunMiller98} for the one degree-of-freedom (quadratic) parabolic barrier.  Calculations for other one degree-of-freedom systems have also been carried out numerically. In particular, the quantum mechanical flux-flux correlation functions for the symmetric and asymmetric Eckart and the double-well potential have been studied in detail (see \cite{CeottoYangMiller05, Miller07}). However, these computations were no {\it ab initio} computations leading to  analytical expressions, but were carried out using various  computational  techniques. Indeed, the development of computational techniques to compute the quantum mechanical flux-flux correlation function has been a subject of great interest since the development of this approach. \todo{In this paragraph I inserted terms like `canonical' and `numerical' to make it more clear what other have done before.}

The traditional approach of quantum mechanics involves  basis function techniques, and a review of this approach is given in \cite{Miller1, Miller98}. While this particular computational approach has proven successful  for ``moderately'' sized molecules, difficulties are encountered when ``large'' systems (such as biomolecules) are considered. The obstacle has been termed  the``exponential wall'' of difficulty that one encounters when attempting  to perform numerical quantum calculations in the traditional manner for many degree of freedom systems \cite{doe2}.  Semiclassical  methods provide an alternative approach. They have been pioneered by Miller and provide significant computational advantages for  ``large'' systems; see \cite{Miller08} for a recent review.

In this paper we present a third approach based on recent work on the phase space approach to quantum mechanics that relies heavily on the dynamical systems framework arising from the application of classical and quantum normal form theory. Classically, this approach has provided a way of constructing a 
recrossing free dividing surface  in the neighborhood of a saddle type equilibrium point which forms the barrier between reactants and products (more precisely the existence of the recrossing free dividing surface is guaranteed for energies not too far away from the energy of the saddle) \cite{wwju,ujpyw,WaalkensBurbanksWigginsb04,WaalkensWiggins04}. The dividing surface at a given energy is bounded by a so called \emph{normally hyperbolic invariant manifold} (NHIM) \cite{Wiggins94} which  as an invariant subsystem with the given energy forms a 'supermolecule' localized between reactants and products which can be viewed as the transition state that gave  TST its name. 
The dividing surface and the NHIM can be constructed in an algorithmic fashion based on a  \emph{classical normal form} (CNF) which can be computed in the neighborhood of the saddle. This is a nonlinear canonical transformation of the phase space coordinates which, to any desired order, decouples the dynamics near the saddle into a single reactive mode and bath modes. (Note: the ``decoupling'' occurs through a local integrable approximation valid in a neighborhood of the saddle, and does in general {\em not} imply separability.)
More recently it has been shown that the classical construction can be generalized to the quantum case \cite{SchubertWaalkensWiggins06,WaalkensSchubertWiggins08,GSWWQuantum}. This essentially involves a systematic quantization of the canonical transformation leading to the decoupling in the classical case. The result is a so called \emph{quantum normal form} (QNF) which leads to a local decoupling also of the quantum dynamics into a reactive mode and bath modes.

The main objective of this paper is to use the classical and quantum normal forms to analytically compute the classical and quantum FFCFs for anharmonic barriers.  The outline  is as  follows. 
In Sec.~\ref{sec:FFCF_NF} we describe how the classical and quantum normal forms can be used to express  the classical and quantum mechanical FFCFs. Sec.~\ref{sec:FFCF_NF}~A is concerned with the classical case, and we show how the dividing surface having the no-recrossing property obtained from classical normal form theory enables a computation of the FFCF that does not require the computation of trajectories. Sec.~\ref{sec:FFCF_NF}~B is concerned with the quantum case and we show how the quantum normal form approach enables a reduction of the computation of the  FFCF in the  general $f$ degree-of-freedom case  to the case of a one degree-of-freedom anharmonic barrier.  Then in  
 Sec.~\ref{sec:1D} we work out in detail the microcanonical quantum FFCF for the example of a barrier in one dimension described by a fourth order anharmonic barrier.
 
 By itself this result should be of some interest because it is the first explicit example of an exact calculation of the FFCF for an anharmonic potential.  
 In Sec.~\ref{sec:concl} we give a discussion of our results and an outlook for future work. The technicalities of two computations are deferred to appendices.

\section{The flux-flux correlation function in the framework of classical and quantum normal form theory}
\label{sec:FFCF_NF}  

We begin by showing how classical and quantum normal form theory can be used to calculate both  
the classical and quantum $\ffcf$ for reactions associated with barrier given by
an index one  saddle of the potential energy surface.  In fact the (slightly more general)
starting point is an equilibrium point of Hamilton's equations which
is of saddle-center-....-center stability type.  For a system with $f$
degrees-of-freedom, this means that the matrix associated with the
linearization of Hamilton's equations about the equilibrium point has
one pair of real eigenvalues $\pm \lambda$ and $f-1$ complex conjugate
pairs of imaginary eigenvalues $\pm \ui \omega_k$, $k=2,\ldots,f$.  We
will call such an equilibrium point a saddle for short.  For
simplicity, we will restrict ourselves to the generic situation where
the linear frequencies $ \omega_k$ are not resonant,
i.e. $m_2\omega_1+\ldots+m_f\omega_f\ne0$ for every nonzero vector of
integers $(m_2,\ldots,m_f)$.

\subsection{The classical case}

Classical normal form (CNF) theory
\cite{ujpyw,WaalkensSchubertWiggins08, GSWWQuantum} provides an algorithm for constructing a (nonlinear)
canonical transformation $\vz\mapsto \vZ = (Q_1, \ldots, Q_f, P_1,\ldots, P_f)$
which, after truncation at a suitable order $N$, leads to an integrable approximation of the dynamics near the saddle.
In terms of the normal form coordinates  $\vZ$
the $N$th order CNF of  the original Hamilton function $H(\vz)$ assumes the following form
\begin{equation} 
\begin{split}
  H_{\mathrm{CNF}}^{(N)} (\vZ) &=  K_{\mathrm{CNF}}^{(N)}
  (I, J_2, J_3, \ldots, J_f)  \\
  &= \sum_{n=0}^{\lfloor N/2 \rfloor} \!  \sum_{|\alpha| = n}
  \kappa_{n,\alpha} I^{\alpha_1} J_2^{\alpha_2} \ldots J_f^{\alpha_f}.
\end{split}
\label{eq:2C-1}
\end{equation}
Here $\lfloor \cdot \rfloor$ denotes the floor function, the $\alpha = ( \alpha_1,\ldots,\alpha_f )$ are vectors with nonnegative integer components and norm
$|\alpha| = \sum_{k=1}^f \alpha_k$,
\begin{equation}
  I = \frac{1}{2} \left( P_1^2 - Q_1^2 \right)
\label{eq:2C-2}
\end{equation}
is an action type integral associated with the reactive mode, and
\begin{equation}
  J_k = \frac{1}{2} \left( P_k^2 + Q_k^2 \right) \,,
  \quad k = 2,\ldots,f \,,
\label{eq:2C-3}
\end{equation}
are action integrals of the bath modes which we group together in the vector $\vJ$. The CNF transformation (including the coefficients
$\kappa_{n,\alpha}$ in \eqref{eq:2C-1}) can be computed in an algorithmic fashion as described in detail in 
\cite{WaalkensSchubertWiggins08}.

In terms of the normal form coordinates the dividing surface is given by
$s(\vZ) = Q_1$. Then,  following the definition 
in Eq.~\eqref{eq:flux_factor_classical} the flux factor is given by
\begin{equation}
\begin{split}
  F (\vZ)= \frac{d}{dt} \Theta\big( {Q_1}_t \big) \bigg|_{t=0} &= - \left\{ H_{\mathrm{CNF}}^{(N)}, \Theta\big( {Q_1} \big)   \right\} (\vZ) \\
   &= \delta(Q_1) \nu(I, \vJ) P_1 \,,
\end{split}
\label{eq:2C-4}
\end{equation}
where 
\begin{equation}
  \nu(I,\vJ) = \frac{\partial}{\partial I} K_{\mathrm{CNF}}^{(N)}(I,\vJ) \,.
\label{eq:2C-5.1}
\end{equation}
Following \eqref{eq:FFCF_classical} the FFCF then takes the form
\begin{align}
  \Ccl(E,t) = \int d\vZ \, &\delta(E-H_{\mathrm{CNF}}^{(N)}(\vZ))
  \nonumber\\ &\times \delta(Q_1) \delta\big( {Q_1}_t \big)
  \nu^2(I,\vJ) P_1 {P_1}_t \,.
\label{eq:2C-5}
\end{align}
The product of the $\delta$-functions of $Q_1$ and the corresponding
time evolved coordinate ${Q_1}_t$ (using the flow associated with $H_{\mathrm{CNF}}^{(N)}$) indicates that only the
infinitesimally short time scales $t \rightarrow 0$ give a
non-vanishing contribution to the integral. The short-time expansion
${Q_1}_t = Q_1 + \nu(I,\vJ) P_1 t + \mathcal{O}(t^2)$ and ${P_1}_t = P_1 +
\mathcal{O}(t)$ yields
\begin{equation} \label{eq:2C-6}
  \Ccl(E,t) = 2 f(E) \delta(t)\,,
\end{equation}
where
\begin{align}
  f(E) &= \frac{1}{2} \int
  d \vZ \, \delta(E-H_{\mathrm{CNF}}^{(N)} (\vZ)) \delta(Q_1) \nu(I,\vJ) |P_1| \nonumber\\
  &= (2\pi)^{f-1} \int_{\mathbb{R}_+^f} dI d\vJ \,
  \delta(E-K_{\mathrm{CNF}}^{(N)}(I,\vJ)) \nu(I,\vJ) \nonumber\\ &= (2\pi)^{f-1}
  \int_{I(E,\vJ)>0} d\vJ \,,
\label{eq:2C-7}
\end{align}
with $I = I(E,\vJ)$ solving the energy equation
$H_{\mathrm{CNF}}^{(N)}(I,\vJ) = E$. The last integral in
Eq.~\eqref{eq:2C-7} is the volume in the space of the center actions
$\vJ= (J_2,\ldots,J_f)$ enclosed by the contour
$H_{\mathrm{CNF}}^{(N)}(0,\vJ) = E$, and accordingly $f(E)$
is nothing but the directional flux through the dividing surface
\cite{WaalkensSchubertWiggins08, GSWWQuantum}. 

We note that formally the result given by Eq.~\eqref{eq:2C-6} exists
in the literature before (e.g., see
Refs.~\cite{Miller98,TrompMiller87} for a corresponding canonical
version of the formula).  However, the contribution of the classical normal form theory in providing a dividing surface with the no-recrossing property  provides a new formula, and interpretation, of the pre-factor $f(E)$  in terms of an integral, in  the bath mode action space, over the NHIM. This eliminates the need to compute trajectories (and the projection function \eqref{eq:projection_operator_classical}) in the computation of the classical FFCF.

\subsection{The quantum mechanical case}
\label{sec:quantumcase}

In  \cite{SchubertWaalkensWiggins06,WaalkensSchubertWiggins08, GSWWQuantum}  a quantum normal
form (QNF) procedure has been developed that yields a local decoupling
of a reactive mode and the bath modes also in the quantum mechanical case if the corresponding classical system has a saddle 
equilibrium of the form described above.
In the quantum case the local simplification of the Hamilton operator is achieved by conjugating it with a suitable unitary transformation.
Similar to the classical case this unitary transformation and the transformed Hamilton operator can be computed in an algorithmic fashion.
The transformed operator then takes the form of a power series 
in terms of elementary operators associated with the reactive and bath modes and, in addition, Planck's constant.
Truncating this expansion at a suitable order $N$ gives the $N$th order QNF approximation
$\hat{H}_{\mathrm{QNF}}^{(N)}$ which is of the form
\begin{align}
  \hat{H}_{\mathrm{QNF}}^{(N)} &= K_{\mathrm{QNF}}^{(N)}(\hat{I},
  \hat{J}_2, \hat{J}_3, \ldots, \hat{J}_f) \nonumber\\
  &= \sum_{n=0}^{\lfloor N/2 \rfloor} \!  \sum_{|\alpha|+j = n}
  \!\!\!\!  \kappa_{n,\alpha,j} \hat{I}^{\alpha_1}
  \hat{J}_2^{\alpha_2} \ldots \hat{J}_f^{\alpha_f} \hbar^j \, .
\label{eq:2Q-1}
\end{align} 
Here, the notation is the same as in \eqref{eq:2C-1}, where in addition 
the $j$ are nonnegative integers and 
\begin{equation}
  \hat{I} = \frac{1}{2} \left( \hat{P}_1^2 - \hat{Q}_1^2 \right)
\label{eq:2Q-2}
\end{equation}
is an operator associated with the reactive mode, and
\begin{equation}
  \hat{J}_k = \frac{1}{2} \left( \hat{P}_k^2 + \hat{Q}_k^2 \right) \,,
  \quad k = 2,\ldots,f \,
\label{eq:2Q-3}
\end{equation}
are operators associated with the bath modes. In  \eqref{eq:2Q-2} and 
\eqref{eq:2Q-3} the $\hat{Q}_k$ and
$\hat{P}_k$, $k = 1,\ldots,f$, are as usual pairs of conjugate position and
momentum operators that satisfy the commutation relations $[\hat{Q}_k,
\hat{Q}_l] = [\hat{P}_k, \hat{P}_l] = 0$ and $[\hat{Q}_k, \hat{P}_l] =
i \hbar \, \delta_{kl}$. The approximation of the original Hamilton operator 
by the QNF in Eq.~\eqref{eq:2Q-1} holds locally in the vicinity of  the saddle equilibrium of the corresponding 
classical system in a sense that is made precise in \cite{WaalkensSchubertWiggins08}.

Since the trace of an operator is invariant under unitary
conjugations of the operator we can evaluate Eq.~\eqref{eq:FFCF_quantum} using the QNF to get
\begin{align}
  &\Cqm(E,t) \nonumber\\ &= \tr \left\{ \delta\left(
      E-\hat{H}_{\mathrm{QNF}}^{(N)} \right)
    \hat{F}_{\mathrm{QNF}}^{(N)} \, e^{i \hat{H}_{\mathrm{QNF}}^{(N)}
      t/ \hbar} \, \hat{F}_{\mathrm{QNF}}^{(N)} \, e^{-i
      \hat{H}_{\mathrm{QNF}}^{(N)} t/ \hbar} \right\}
\label{eq:2Q-8}
\end{align}
with the flux operator given by
\begin{equation}
  \hat{F}_{\mathrm{QNF}}^{(N)} = \frac{i}{\hbar}
  \big[ \hat{H}_{\mathrm{QNF}}^{(N)}, \Theta(\hat{Q}_1) \big] \,.
\label{eq:2Q-9}
\end{equation}

Since the operators $\hat{I}$ and $\hat{J}_k$, $k=2,\ldots,f$, mutually commute
 the eigenstates of $\hat{H}_{\mathrm{QNF}}^{(N)}$ can
be chosen such that they are simultaneously the  eigenstates of all the elementary operators
$\hat{I}$ and $\hat{J}_k$, whose spectral properties are well
known. Thus,
\begin{equation}
  \hat{H}_{\mathrm{QNF}}^{(N)} |I,n_2,\ldots,n_f\rangle = E |I,n_2,\ldots,n_f\rangle
\label{eq:2Q-4}
\end{equation}
with
\begin{equation}
  |I,n_2,\ldots,n_f\rangle = |\psi_I\rangle \otimes |\psi_{n_2}\rangle
  \otimes \ldots \otimes |\psi_{n_f}\rangle \,,
\label{eq:2Q-5}
\end{equation}
where
\begin{subequations}
\begin{alignat}{3}
  &\hat{I} |\psi_I\rangle = I |\psi_I\rangle \, ,& \qquad  &I \in \mathbb{R} \, , \label{eq:2Q-6.a}\\
  &\hat{J}_k |\psi_{n_k}\rangle = \hbar (n_k + 1/2) |\psi_{n_k}\rangle
  \, ,& \qquad &n_k \in \mathbb{N}_0 \, , \label{eq:2Q-6.b}
\end{alignat}
\label{eq:2Q-6}
\end{subequations}
and
\begin{equation}
  E = K_{\mathrm{QNF}}^{(N)} \big( I, \hbar (n_2+1/2), \ldots, \hbar (n_f+1/2) \big) \, .
\label{eq:2Q-7}
\end{equation}

Using the basis given by the eigenstates $|I,n_2,\ldots,n_f\rangle$
one can now straightforwardly trace out the bath modes in
Eq.~\eqref{eq:2Q-8}. Indeed, let us define the operator
\begin{align}
  \hat{\mathcal{H}}&_{n_2,\ldots,n_f}^{(N)} = \sum_{n=0}^{\lfloor N/2
    \rfloor} \!  \sum_{|\alpha|+j = n}
  \!\!\!\!  \kappa_{n,\alpha,j} \hat{I}^{\alpha_1} \nonumber\\
  &\times \left( n_2+\frac{1}{2} \right)^{\alpha_2} \ldots \left(
    n_f+\frac{1}{2} \right)^{\alpha_f} \hbar^{|\alpha|-\alpha_1+j}
\label{eq:2Q-10}
\end{align}
parametrized by the $(f-1)$ nonnegative quantum numbers $n_2$, \ldots,
$n_f$. Then Eqs~\eqref{eq:2Q-8} and \eqref{eq:2Q-9} can be
written as
\begin{equation}
  \Cqm(E,t) = \sum_{n_2,\ldots,n_f} \tr \left\{ \delta(
    E-\hat{\mathcal{H}}) \hat{\mathcal{F}} e^{i \hat{\mathcal{H}} t/
      \hbar} \hat{\mathcal{F}} e^{-i \hat{\mathcal{H}} t/ \hbar}
  \right\}
\label{eq:2Q-11}
\end{equation}
and
\begin{equation}
  \hat{\mathcal{F}} = \frac{i}{\hbar}
  \big[ \hat{\mathcal{H}}, \Theta(\hat{Q}_1) \big]
\label{eq:2Q-12}
\end{equation}
respectively, where, to avoid a cumbersome notation,  we have dropped the
superscript $N$ and subscripts $n_2,\ldots,n_f$ for the operators $\hat{\mathcal{H}}$ and $\hat{\mathcal{F}} $.

Equations~\eqref{eq:2Q-11} and \eqref{eq:2Q-12} show that the problem
of calculating the quantum $\ffcf$ for a system with $f>1$ degrees of
freedom effectively reduces to the corresponding problem for a
one-dimensional system described by a Hamiltonian
of the form Eq.~\eqref{eq:2Q-10}
which is a polynomial of the operator $\hat{I}$ associated with the reactive mode only.
In the following section
we present an explicit calculation of the $\ffcf$ for the simplest
{\it anharmonic} one-dimensional Hamiltonian of this form.

\section{The quantum flux-flux correlation function for one dimensional anharmonic barriers}
\label{sec:1D}

We now present an analytical calculation of the $\ffcf$ for the
two-parameter 
family of Hamilton operators defined by
\begin{equation}
  \hat{H}(a, \lambda) = \hat{h} + a \hat{h}^2 \,,
  \quad \hat{h} = \frac{1}{2} \left( \hat{p}^2 - \lambda^2 \hat{q}^2 \right) \,.
\label{eq:3-01}
\end{equation} 
Here $\lambda$ parametrizes the width of the barrier in the harmonic approximation and $a$ characterizes the anharmonicity of the barrier.
The Hamiltonian operator $\hat{H}$ can be viewed  to be in  
quantum normal form. In fact the operator $\hat{h}$ differs from the operator  $\hat{I}$ defined in Sec.~\ref{sec:quantumcase} only by a factor of $\lambda$ which follows from a linear transformation of 
$\hat{p}$ and $\hat{q}$ which does not alter the normal form procedure described in the previous section.  The Hamilton operator in \eqref{eq:3-01} can therefore be 
considered to describe the simplest possible anharmonic barrier. 

The starting point of our calculation is the system of eigenstates of
$\hat{h}$, (and therefore of $\hat{H}$) 
\begin{equation}
  \hat{h} | \psi_{\sigma}^E \ra = E | \psi_{\sigma}^E \ra \,, 
  \quad E \in \mathbb{R} \,, \quad \sigma = \pm 1\,.
\label{eq:3-02}
\end{equation}
The corresponding wavefunctions are \cite{Chruscinski1,  Chruscinski2}
\begin{align}
  \la q | \psi_{\sigma}^E \ra = & \, \frac{1}{2\pi\hbar} \left(
    \frac{2\hbar}{\lambda} \right)^{\frac{1}{4}} \exp\left(
    {\frac{\pi}{4} \frac{E}{\lambda\hbar}} \right)
  \Gamma\left( \frac{1}{2}-i\frac{E}{\lambda\hbar} \right) \nonumber\\
  &\times D_{-\frac{1}{2}+i\frac{E}{\lambda\hbar}}\left( \sigma
    e^{-i\frac{\pi}{4}} \sqrt{\frac{2\lambda}{\hbar}} \, q \right) \,,
\label{eq:3-03}
\end{align}
where $D_\nu$ denotes the parabolic cylinder function of order $\nu$
\cite{AbramovizStegun}. The eigenstates are mutually orthogonal,
\begin{equation}
  \la \psi_{\sigma}^E | \psi_{\sigma'}^{E'} \ra = \delta_{\sigma,\sigma'} \, \delta(E-E') \,,
\label{eq:3-04}
\end{equation}
and form a complete basis, i.e.,
\begin{equation}
  \sum_{\sigma=\pm 1} \int_{-\infty}^{+\infty} dE \: | \psi_{\sigma}^E \ra \la \psi_{\sigma}^E | = \hat{\mathbf{1}} \,,
\label{eq:3-05}
\end{equation}
where $\hat{\mathbf{1}}$ denotes the identity operator.

Using the $\psi_{\sigma}^E$ basis to expand the trace in
Eq.~\eqref{eq:2Q-11} we can write the quantum $\ffcf$ as
\begin{align}
  &\Cqm(E,t,a) = \iint dE' dE'' \: \delta \left(E''+aE''^2-E\right) \nonumber\\
  &\times \exp\left[ \frac{i t}{\hbar}\left(E'+aE'^2-E\right) \right] \,
  \sum_{\sigma,\sigma'} \left| \la \psi_{\sigma}^{E''} | \hat{F} |
    \psi_{\sigma'}^{E'} \ra \right|^2 \,,
\label{eq:3-06}
\end{align}
where for the discussion below,  we explicitly added the anharmonicity parameter $a$ to the
argument of the $\ffcf$ (note that as opposed to $a$ the parameter $\lambda$ can in principle be removed by a suitable scaling of the energy). 
If we denote the two solutions of
$\tilde{E}+a\tilde{E}^2=E$ by
\begin{equation}
  \tilde{E}_{\sigma} = \frac{1}{2a} \left( -1 + \sigma \sqrt{1+4aE} \right) ,
  \quad \sigma = \pm 1 \,,
\label{eq:3-07}
\end{equation}
then \eqref{eq:3-06} becomes
\begin{align}
  \Cqm(E,t,a) = \; &\frac{1}{\sqrt{1+4aE}} \int_{-\infty}^{+\infty}
  dE' \, e^{i (E'+aE'^2-E) t/\hbar} \nonumber\\ &\times
  \sum_{\sigma,\sigma',\sigma''} \left| \la
    \psi_{\sigma}^{\tilde{E}_{\sigma''}} | \hat{F} |
    \psi_{\sigma'}^{E'} \ra \right|^2
\label{eq:3-08}
\end{align}
if   $1+4aE>0$, and $\Cqm(E,t,a) = 0$ else.
The latter condition on the energy $E$ and the parameter $a$ simply assures
that we actually have a barrier scattering problem if the inequality  is satisfied.

The matrix elements of the flux operator $\hat{F}$ are calculated as
follows. According to Eq.~\eqref{eq:2Q-12} we have
\begin{equation}
  \hat{F} = \frac{i}{\hbar} \big[ \hat{h}+a\hat{h}^2, \Theta(\hat{q}) \big] \,,
\label{eq:3-09}
\end{equation}
so that
\begin{align}
  \la \psi_{\sigma}^E | \hat{F} | \psi_{\sigma'}^{E'} \ra = &
  \frac{i}{\hbar} \left[ (E+aE^2) - (E'+aE'^2) \right] \nonumber\\
  &\times \int_0^{\infty} dq \, \la \psi_{\sigma}^E | q \ra \la q |
  \psi_{\sigma'}^{E'} \ra \,.
\label{eq:3-10}
\end{align}
Then, using Eq.~\eqref{eq:3-03} into Eq.~\eqref{eq:3-10} and
performing the integration over $q$ we obtain
\begin{align}
  \la \psi_{\sigma}^E &| \hat{F} | \psi_{\sigma'}^{E'} \ra =
  \frac{1+a(E+E')}{8\pi^2\hbar} e^{\frac{\pi}{4} \frac{E+E'}{\lambda
      \hbar}} 2^{\frac{i}{2} \frac{E-E'}{\lambda \hbar}} \nonumber\\
  &\times \Bigg\{ \sigma e^{-i \frac{\pi}{4}} \, \Gamma\left(
    \frac{3}{4}+i\frac{E}{2\lambda\hbar} \right) \Gamma\left(
    \frac{1}{4}-i\frac{E'}{2\lambda\hbar} \right) \nonumber\\
  &\phantom{xxx} + \sigma' e^{i \frac{\pi}{4}} \, \Gamma\left(
    \frac{1}{4}+i\frac{E}{2\lambda\hbar} \right) \Gamma\left(
    \frac{3}{4}-i\frac{E'}{2\lambda\hbar} \right) \Bigg\} \,.
\label{eq:3-11}
\end{align}
This finally leads to the following expression for the double sum
(over $\sigma$ and $\sigma'$) entering Eqs.~\eqref{eq:3-06} and
\eqref{eq:3-08}:
\begin{align}
  \sum_{\sigma,\sigma'} &\left| \la \psi_{\sigma}^E | \hat{F} |
    \psi_{\sigma'}^{E'} \ra \right|^2 = \frac{\left[ 1 + a (E+E')
    \right]^2}{16 \pi^4 \hbar^2} \exp \left( \frac{\pi}{2}
    \frac{E+E'}{\lambda \hbar} \right) \nonumber\\ &\times \Bigg\{
  \left| \Gamma \left( \frac{1}{4} + i \frac{E}{2\lambda\hbar} \right)
  \right|^2 \left| \Gamma \left( \frac{3}{4} + i
      \frac{E'}{2\lambda\hbar} \right) \right|^2 \nonumber\\
  &\phantom{xxx} + \left| \Gamma \left( \frac{3}{4} + i
      \frac{E}{2\lambda\hbar} \right) \right|^2 \left| \Gamma \left(
      \frac{1}{4} + i \frac{E'}{2\lambda\hbar} \right) \right|^2
  \Bigg\} \,.
\label{eq:3-12}
\end{align}

We then substitute Eq.~\eqref{eq:3-12} into Eq.~\eqref{eq:3-08} and
use the formula (see Appendix~\ref{app:integral} for its derivation)
\begin{align}
  \int_{-\infty}^{+\infty} &dx \, e^{i A x^2 + B x} (1+Cx)^2 \, \big|
  \Gamma(D+ix) \big|^2 \nonumber\\ &= 2\pi \frac{\Gamma(2D)}{2^{2D}}
  \exp \left( -i \frac{A}{4} \frac{\partial^2}{\partial x^2} \right)
  \nonumber\\ &\phantom{xx} \frac{(\cosh x + i CD \sinh
    x)^2+\frac{1}{2}C^2D}{(\cosh x)^{2D+2}} \Bigg|_{x=-i\frac{B}{2}}
\label{eq:3-13}
\end{align}
with $A = 4a \lambda^2 \hbar t$, $B = \pi + 2i \lambda t$, $C =
2\lambda t a / (1+a\tilde{E})$, and $D = 1/4$ or $3/4$ to arrive at
the central result of our paper:
\begin{subequations}
\begin{equation}
  \Cqm(E,t,a) = \frac{\lambda e^{i\pi/4}}{2^{9/2} \pi^{5/2} \hbar} \,
  \Lambda \big( E/\lambda\hbar, \lambda t, a \lambda\hbar \big) \,,
\label{eq:3-14}
\end{equation}
where
\begin{align}
  \Lambda(\varepsilon,\tau,\alpha) =
  &\frac{e^{-i\varepsilon\tau}}{\sqrt{1+4\varepsilon\alpha}}
  \sum_{\sigma = \pm 1} \exp\left( \frac{\pi \epsilon_{\sigma}}{2}
  \right) \Bigg\{ \nonumber\\ &4 \,\bigg| \Gamma \left( \frac{3}{4} +
    i \frac{\epsilon_{\sigma}}{2} \right) \bigg|^2 \Omega_{-\sigma}
  \left( \frac{1}{4}; \varepsilon,\tau,\alpha \right) \nonumber\\ &+ i
  \, \bigg| \Gamma \left( \frac{1}{4} + i \,
    \frac{\epsilon_{\sigma}}{2} \right) \bigg|^2 \Omega_{-\sigma}
  \left( \frac{3}{4}; \varepsilon,\tau,\alpha \right) \Bigg\} \,,
\label{eq:3-15}
\end{align}
\begin{equation}
  \epsilon_{\sigma}(\varepsilon,\alpha) = \frac{1}{2\alpha} 
  \left( -1 + \sigma \sqrt{1+4\varepsilon\alpha} \right) ,
\label{eq:3-16}
\end{equation}
and 
\begin{align}
  \Omega&_{\sigma}(\nu;\varepsilon,\tau,\alpha) \nonumber\\
  &= \alpha^2 \exp\left( -i\alpha\tau
    \frac{\partial^2}{\partial\tau^2} \right) \frac{(\epsilon_{\sigma}
    \sinh\tau - 2i \nu \cosh\tau)^2-2\nu}{(\sinh\tau)^{2\nu+2}} \,.
\label{eq:3-17}
\end{align}
\label{eq:main_result}
\end{subequations}
The expression for the $\ffcf$ given by Eq.~\eqref{eq:main_result} is
{\it exact} and holds for all energies $E$ and parameters $a$ satisfying $1+4aE>0$. (As shown
above, $\Cqm(E,t,a) = 0$ if $1+4aE < 0$.) 
In the following we consider the limits of an harmonic saddle, $a\to 0$,
and short times $\lambda t \ll 1$ for which cases $\Omega_{\sigma}$ in Eq.~\eqref{eq:3-17}  
assumes a simpler form.


\subsection{The case of a harmonic barrier  ($a=0$)}

In the limit $\alpha \rightarrow 0$ Eq.~\eqref{eq:3-16} yields
$\epsilon_1 = \varepsilon + \mathcal{O}(\alpha)$ and $\epsilon_{-1}
= -1/\alpha - \varepsilon + \mathcal{O}(\alpha)$. It is then
straightforward to show that for $\alpha = 0$ the $\sigma=-1$
contribution to the sum in the right-hand side of Eq.~\eqref{eq:3-15}
vanishes and thus 
\begin{equation}
  \Lambda(\varepsilon,\tau,0) = e^{\varepsilon \left( 
      \frac{\pi}{2}-i\tau \right)} \left[ 4\frac{\left| \Gamma\left( 
          \frac{3}{4}+\frac{i\varepsilon}{2} \right) \right|^2}{(\sinh\tau)^{1/2}} + i\frac{\left| \Gamma\left( 
          \frac{1}{4}+\frac{i\varepsilon}{2} \right) \right|^2}{(\sinh\tau)^{3/2}}  \right] ,
\label{eq:3-18}
\end{equation}
which is valid for all energies, $\varepsilon \in \mathbb{R}$.

\begin{figure}[t]
\centerline{\epsfig{figure=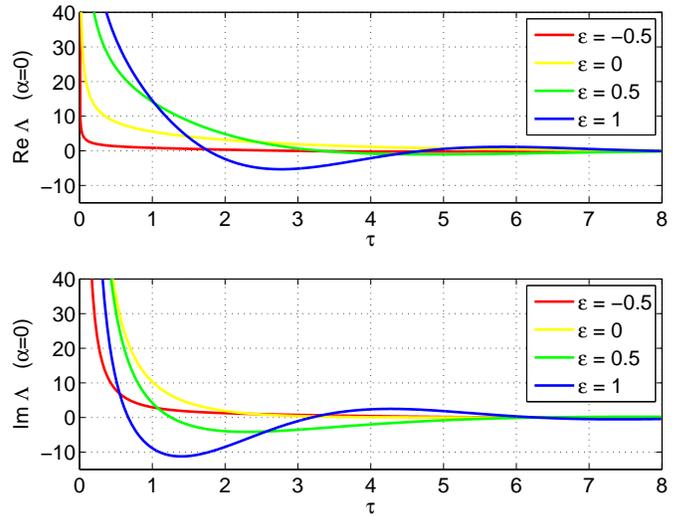,width=3.5in}}
\caption{(Color online) The time dependence of the real (top) and imaginary (bottom) parts of the function $\Lambda$ in \eqref{eq:3-18} which enters the quantum $\ffcf$ for an harmonic barrier in 
\eqref{eq:3-14}. The graphs are for different values of the scaled energy $\epsilon$.}
\label{fig1}
\end{figure}

Equations~\eqref{eq:3-14} and \eqref{eq:3-18} yield an exact
expression for the {\it microcanonical} quantum $\ffcf$ of the parabolic
barrier system with Hamiltonian
$(\hat{p}^2-\lambda^2 \hat{q}^2)/2$. The corresponding {\it canonical}
version of the $\ffcf$, defined as
\begin{equation}
  \Ccan(\beta,t,a) = \tr \left\{ e^{-\beta \hat{H}} \hat{F}
    e^{i \hat{H} t/ \hbar} \hat{F} e^{-i \hat{H} t/ \hbar} \right\} \,,
\label{eq:3-18.1}
\end{equation}
can be readily calculated for the case of $a=0$ by performing the
bilateral Laplace transformation of \eqref{eq:3-14} which gives
\begin{align}
  \Ccan(\beta,t,0) &= \int_{-\infty}^{+\infty} dE \, e^{-\beta E}
  \Cqm(E,t,0) \nonumber\\ &= \frac{\lambda^2}{4\pi}
  \frac{\cosh(\lambda t_c) \sinh(\lambda\hbar\beta/2)}{\left[
      \sinh^2(\lambda t_c)+\cosh^2(\lambda\hbar\beta/2) \right]^{3/2}}
\label{eq:3-18.2}
\end{align}
with $t_c = t-i\hbar\beta/2$. Here Eq.~\eqref{eq:3-13} with $A=C=0$ was
used to calculate the integral over energy.

We note that Eq.~\eqref{eq:3-18.2} was originally obtained by
Miller {\it et al.}  \cite{MillerSchwartzTromp83} by representing the
$\ffcf$ in terms of the time evolution operator for the harmonic
oscillator. However, to our
knowledge, the explicit expression for the microcanonical $\ffcf$ in
Eqs.~\eqref{eq:3-14} and \eqref{eq:3-18} has not  been reported in the literature before.

In Fig.~\ref{fig1} the time dependence of the function $\Lambda$ in \eqref{eq:3-18} is shown for different values of the scaled energy $\epsilon$. We note that $\Lambda$ and hence the microcanonical $\ffcf$ always diverge in the limit $t\to 0$ and depending on the energy shows more or less pronounced oscillations.


\subsection{The short-time regime  ($\lambda t \ll 1$)}
\label{sec:short-timeregime}

For short times, $\tau \ll 1$, one can approximate the hyperbolic
functions on the right-hand side of Eq.~\eqref{eq:3-17} by their
leading order Taylor expansions to obtain
\begin{align}
  \Omega&_{\sigma}(\nu;\varepsilon,\tau,\alpha) \nonumber\\ &\simeq
  \alpha^2 \big( \epsilon_{\sigma}^2 \mathcal{D}_{2\nu} - 4i \nu
  \epsilon_{\sigma} \mathcal{D}_{2\nu+1} + 2\nu(2\nu+1)
  \mathcal{D}_{2\nu+2} \big)
\label{eq:3-19}
\end{align}
with
\begin{equation}
  \mathcal{D}_\mu(\tau,\alpha) = \exp\left( -i\alpha\tau
    \frac{\partial^2}{\partial\tau^2} \right) \frac{1}{\tau^\mu} \,.
\label{eq:3-20}
\end{equation}
As shown in Appendix~\ref{app:exp_deriv} $\mathcal{D}_\mu$ in Eq.~\eqref{eq:3-20} can be written as
\begin{equation}
  \mathcal{D}_\mu(\tau,\alpha) = e^{\gamma^2 \! / 4} \, D_{-\mu}(\gamma) 
  \left(\frac{\gamma}{\tau}\right)^\mu
  , \quad \gamma = \left( \frac{\tau}{2\alpha} e^{-i\pi/2} \right)^{1/2} \!.
\label{eq:3-21}
\end{equation}
Equations~(\ref{eq:3-14}-\ref{eq:3-16}) together with
Eqs.~\eqref{eq:3-19} and \eqref{eq:3-21} provide an explicit
expression for the $\ffcf$ at short times.

\begin{figure}[t]
\centerline{\epsfig{figure=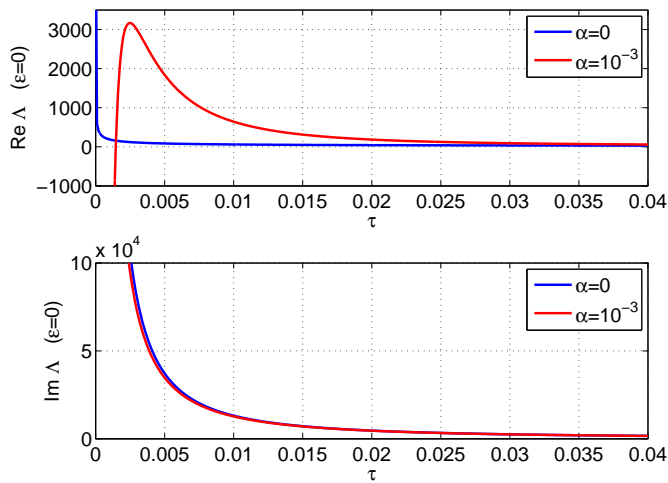,width=3.5in}}
\caption{(Color online) Short time regime of the real (top) and imaginary (bottom) parts of the function $\Lambda$ which enters the quantum $\ffcf$  in \eqref{eq:3-14} (see  Sec.~\ref{sec:short-timeregime}). The graphs are for different values of the scaled anharmonicity parameter $\alpha$ and all have energy zero. }
\label{fig2}
\end{figure}

Figure~\ref{fig2} compares the time decay of the dimensionless
correlation function $\Lambda$ in the harmonic case ($\alpha=0$, blue
lines), given by Eq.~\eqref{eq:3-18}, and that in the anharmonic case
($\alpha=10^{-3}$, red lines), given by Eqs.~\eqref{eq:3-15},
\eqref{eq:3-19}, and \eqref{eq:3-21}. One sees that even for very
small (but non-vanishing) values of the dimensionless anharmonicity
parameter $\alpha$ the time dependence of the $\ffcf$ significantly
differs from that of the corresponding harmonic problem. In fact, it
is straightforward to show that as $\tau \rightarrow 0$ one has
$\mathrm{Re} \, \Lambda \rightarrow +\infty$ for $\alpha=0$, while
$\mathrm{Re} \, \Lambda \rightarrow -\infty $ for $\alpha>0$. The
transition from the anharmonic case to the harmonic one takes place in
a discontinuous manner: as $\alpha$ tends to zero the maximum of
$\mathrm{Re} \, \Lambda$ (i.e., the peak of the red curve in the upper
half of Fig.~\ref{fig2}) becomes higher and sharper and approaches
$\tau=0$ recovering the harmonic result (the monotonic blue curve in
Fig.~\ref{fig2}).

\section{Conclusions}
\label{sec:concl}

In this paper we have presented a method for computing classical and quantum
flux-flux correlation functions for reactive systems with a potential
barrier characterized by saddle type equilibria in phase space. The
method is based on the normal form transformation of the system's
Hamiltonian in a vicinity of the saddle point.

In the classical case, the time dependence of the correlation function
(with respect to a recrossing free dividing surface) is given by the
 $\delta$-function.  While this form of the time dependence  has been known in the  theoretical chemistry community for some time, the contribution of the classical normal form theory is that it provides a dividing surface having the no-recrossing property that allows the computation of the pre-factor (essentially the flux through the dividing surface). No computation of trajectories is required to evaluate the flux-flux correlation function. The time integration to compute the rate becomes trivial. 
 
In the quantum case, we showed that the problem of calculating the
correlation function in a system with more than one degree-of-freedom
reduces to an effective one degree-of-freedom problem. The Hamiltonian
of this effective one degree-of-freedom system is obtained through the
quantum normal form procedure. Finally, and most importantly, we
derive (for the first time in the literature) an analytical expression
for the flux-flux correlation function for the simplest {\it
  anharmonic} one-dimensional Hamiltonian in quantum normal form.

\acknowledgments

A.G. and H.W. acknowledge support by EPSRC under grant
No. EP/E024629/1.  S.W.  acknowledges the support of the  Office of Naval Research Grant No.~N00014-01-1-0769.

\appendix
\section{Derivation of Eq.~(\ref{eq:3-13})}
\label{app:integral}

We begin our derivation of the formula, Eq.~\eqref{eq:3-13}, by
writing
\begin{equation}
  \big| \Gamma(D+ix) \big|^2 = \frac{\Gamma(2D)}{2^{2D}} 
  \int_{-\infty}^{+\infty} dq \, \frac{e^{-iqx}}{\big[ \cosh(q/2) \big]^{2D}} \,.
\label{eq:app1-1}
\end{equation}
The integral representation given by Eq.~\eqref{eq:app1-1} is readily
obtained, e.g., from formula 5.13.2 in Ref.~\cite{nist10}. Then,
denoting the left-hand side of Eq.~\eqref{eq:3-13} by $\mathcal{I}$ we
get
\begin{align}
  \mathcal{I} = &\frac{\Gamma(2D)}{2^{2D}} \int_{-\infty}^{+\infty}
  \frac{dq}{\big[ \cosh(q/2) \big]^{2D}} \nonumber\\
  &\times \int_{-\infty}^{+\infty} dx (1+Cx)^2 e^{iAx^2-i(q+iB)x} \,.
\label{eq:app1-2}
\end{align}
The second integral in the right-hand side of Eq.~\eqref{eq:app1-2}
can be written as
\begin{equation}
  \sum_{n=0}^\infty \frac{(iA)^n}{n!} \int_{-\infty}^{+\infty} dx \,
   x^{2n} (1+Cx)^2 e^{-i(q+iB)x} \,.
\label{eq:app1-3}
\end{equation}
Now, using
\begin{equation}
  \int_{-\infty}^{+\infty} dx x^n e^{-i(q+iB)x} = 2\pi i^n \delta^{(n)}(q+iB) \,,
\label{eq:app1-4}
\end{equation}
with $\delta^{(n)}$ denoting the $n$th derivative of the delta
function, and then, carrying out the $q$-integration in
Eq.~\eqref{eq:app1-3} we obtain
\begin{align}
  \mathcal{I} = 2\pi &\frac{\Gamma(2D)}{2^{2D}} \sum_{n=0}^\infty
  \frac{(-iA)^n}{n!} \frac{\partial^{2n}}{\partial q^{2n}} \nonumber\\
  & \frac{\big( \cosh\frac{q}{2}+iCD\sinh\frac{q}{2} \big)^2+\frac{1}{2}C^2D}{\big(
    \cosh\frac{q}{2} \big)^{2D+2}} \Bigg|_{q=-iB} \!\!\! .
\label{eq:app1-5}
\end{align}
Finally, formally summing the series,
\begin{equation}
  \sum_{n=0}^\infty \frac{(-iA)^n}{n!} \frac{\partial^{2n}}{\partial q^{2n}} \, f(q)
  = \exp\left( -iA \frac{\partial^2}{\partial q^2} \right) f(q)
\label{eq:app1-6}
\end{equation}
with $f$ denoting an arbitrary function, and making the change $x = q/2$
we arrive at Eq.~\eqref{eq:3-13}.


\section{Derivation of Eq.~(\ref{eq:3-21})}
\label{app:exp_deriv}

Taking into account the identity
\begin{equation}
  \frac{1}{\tau^\mu} = \frac{1}{\Gamma(\mu)} \int_0^\infty ds \, s^{\mu-1} e^{-\tau s}
\label{eq:app2-1}
\end{equation}
we rewrite Eq.~\eqref{eq:3-20} as
\begin{equation}
  \mathcal{D}_\mu = \frac{1}{\Gamma(\mu)} \int_0^\infty ds \, s^{\mu-1} e^{-i\alpha\tau s^2 - \tau s} \,.
\label{eq:app2-2}
\end{equation}
Then, using formula 3.462.1 in Ref.~\cite{GR00Table},
\begin{align}
  \int_0^\infty &ds \, s^{\mu-1} e^{-A s^2 - B s} \nonumber\\ &=
  \frac{\Gamma(\mu)}{(2A)^{\mu/2}} \exp\left( \frac{B^2}{8A} \right)
  D_{-\mu}\left( \frac{B}{\sqrt{2B}}\right) \,,
\label{eq:app2-3}
\end{align}
for $A = e^{i\pi/2} \alpha\tau$ and $B = \tau$ we arrive at
Eq.~\eqref{eq:3-21}.

\bibliographystyle{aipnum4-1}
\def\cprime{$'$}

\end{document}